# Role of entropy in the thermodynamic evolution of the time scale of molecular dynamics near the glass transition

K. Grzybowska,[1,2,*] A. Grzybowski,[1,2,#] S. Pawlus,[1,2] J. Pionteck,[3] and M. Paluch,[1,2]

[1]*Institute of Physics, University of Silesia, Uniwersytecka 4, 40-007 Katowice, Poland*

[2]*Silesian Center for Education and Interdisciplinary Research, 75 Pułku Piechoty 1A, 41-500 Chorzów, Poland*

[3]*Leibniz Institute of Polymer Research Dresden, Hohe Str. 6, D-01069 Dresden, Germany*

[*] Corresponding author's email: katarzyna.grzybowska@us.edu.pl
[#] Corresponding author's email: andrzej.grzybowski@us.edu.pl

ABSTRACT

In this Letter, we investigate how changes in the system entropy influence the characteristic time scale of the system molecular dynamics near the glass transition. Independently of any model of thermodynamic evolution of the time scale, against some previous suppositions, we show that the system entropy $S$ is not sufficient to govern the time scale defined by structural relaxation time $\tau$. In the density scaling regime, we argue that the decoupling between $\tau$ and $S$ is a consequence of different values of the scaling exponents $\gamma$ and $\gamma_S$ in the density scaling laws, $\tau=f(\rho^{\gamma}/T)$ and $S=h(\rho^{\gamma_S}/T)$, where $\rho$ and $T$ denote density and temperature, respectively. It implies that the proper relation between $\tau$ and $S$ requires supplementing with a density factor, $u(\rho)$, i.e., $\tau=g(u(\rho)w(S))$. This meaningful finding additionally demonstrates that the density scaling idea can be successfully used to separate physically relevant contributions to the time scale of molecular dynamics near the glass transition. As an example, we revise the Avramov entropic model of the dependence $\tau(T,\rho)$, giving evidence that its entropic basis has to be extended by the density dependence of the maximal energy barrier for structural relaxation. We also discuss the excess entropy $S_{ex}$, the density scaling of which is found to mimic the density scaling of the total system entropy $S$.

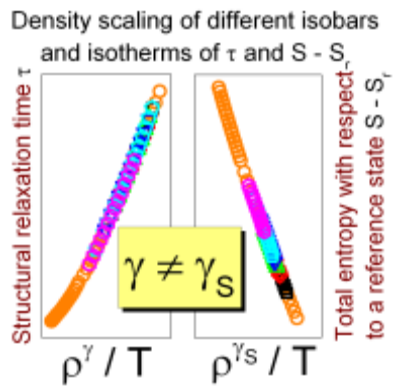

Despite several decades of studies, our understanding of the glass transition and related phenomena is still unsatisfactory. For instance, a fervent debate on relevant factors, which govern the dramatic slowdown in molecular dynamics of materials on approaching the glassy state, has not resulted yet in any complete and commonly accepted theoretical explanation of the intriguing dynamic behavior. Some first attempts[1,2,3,4] at modelling the temperature evolution of the time scale of molecular dynamics near the glass transition, which are based separately on either the pure thermal activation of molecular dynamics or the pure free volume effect on molecular mobility, have turned out not to be well grounded although they provided a satisfactory phenomenological description of the temperature dependence of structural relaxation times $\tau(T)$. This finding has been confirmed by studying the ratios of the activation energies $E_V/E_p$ determined respectively at a constant volume $V$ and a constant pressure $p$ from experimental data measured at ambient and high pressures for various glass formers from different material groups, which have clearly shown that both the density and temperature fluctuations contribute to molecular dynamics near the glass transition (e.g, $E_V/E_p \approx 0.5$ for van der Waals supercooled liquids, which means that the thermal activation and the free volume play comparable roles in molecular dynamics of such materials).[5,6]

A leading trend in exploring the physicochemical phenomena that occur near the glass transition invokes the dominant role of entropy in controlling molecular dynamics. In fact, a proper description of the effect of temperature-density fluctuations on the molecular behavior of glass formers can be seen in the entropic models of the thermodynamic evolution of the dynamic time scale. Among such models, the historically most prominent one is that formulated by Adam and Gibbs (AG) a half-century ago.[7] The AG model constituted a paradigm of dynamic heterogeneity in the glass transition physics by the assumption of the intermolecular cooperative rearrangement regions (CRR), the size of which increases while the system is approaching the glassy state. The related decrease in the configurational entropy of the system, $S_c$, results in the rapid growth of the time scale that depends on the product of $T$ and $S_c$ as folows $\tau=\tau_0\exp(A_G/TS_c)$, where $A_G$ and $\tau_0$ are fitting parameters. Although a lot of previous valuable studies have been based on the AG approach, this model has difficulty in analyzing experimental data measured under high pressure, starting from technical inconveniences due to many parameter complex extensions of the AG temperature model to its temperature-pressure versions[8,9,10,11] and ending with a serious doubt about its theoretical validity in context of the recently observed density scaling of molecular dynamics of viscous liquids.

In the last decade, a significant progress in developing the glass transition physics has been made by the phenomenological observation that experimental dynamic data measured in isobaric and isothermal conditions (such as structural relaxation times and viscosities) can be plotted onto a master curve as a function of a single variable $\rho^\gamma/T$, where $\gamma$ is a material constant. It has turned out that the power law density scaling law, $\tau = f(\rho^\gamma/T)$, is obeyed by many glass formers that belong to different material groups such as van der Waals supercooled liquids, polymer melts, supercooled ionic liquids, and liquid crystals.[5,6] What is more, the scaling exponent $\gamma$ has been related to the exponent of the repulsive part of the effective short-range intermolecular potential based on the simple Lennard-Jones potential.[12,13,14,15,16,17,18,19] Later, a solid theoretical ground for the density scaling idea has been formulated by Dyre's group in the theory of isomorphs,[20] which provides rules that should be obeyed to ensure that molecular dynamics, thermodynamics and structure of *a simple liquid* are isomorph invariant (i.e., invariant along an isomorph that is a curve in the $(\rho,T)$ phase diagram, which can be in practice found from the general density scaling criterion,[12,21,22] $\tilde{h}(\rho)/T = const$ at a constant $\tau$ expressed in reduced units defined by the theory of isomorphs, where $\tilde{h}(\rho) = \rho^\gamma$ in most experimental cases until recently known). The theory of isomorphs implies that both the configurational entropy $S_c$ and the structural relaxation time $\tau$ should be isomorph invariant at least to a good approximation. However, the product $TS_c$, which constitutes a single variable in the AG model, is not isomorph invariant. It means that the original AG model does not comply with the theory of isomorphs and is not appropriate to describe the density scaling of molecular dynamics near the glass transition. Since the dynamic scaling idea has turned out to be very attractive, some assumptions have been recently suggested[12,23,24] to adjust the AG model to this idea. Nevertheless, the density scaling of molecular dynamics is not inherent in the original AG model.

The recent achievements based on the density scaling idea demand of us that the models valid to describe the thermodynamic evolution of the time scale of molecular dynamics should obey the density scaling law. In this Letter, we thoroughly analyze such a model, which enables us to discuss a proper relation between the time scale of molecular dynamics and the total system entropy in the density scaling regime.

The first well-acknowledged model,[6] which is morphologically ready to reproduce the power law density scaling, has been formulated by Casalini et al.[25]

$$\tau = \tau_0 \exp\left[\left(A\rho^\gamma / T\right)^D\right] \qquad (1)$$

where $\tau_0$, $A$, $D$, and $\gamma$ are fitting parameters and the density $\rho$ can be replaced with the inverse specific volume $V^{-1}$. Thus, the value of the scaling exponent $\gamma$ can be easily found from fitting experimental dependences $\tau(T,\rho)$ or $\tau(T,V)$ to Eq. (1). This temperature-density model is based on the Avramov temperature-pressure model[26] that assumes that the structural relaxation time is inversely proportional to an average frequency, $<\nu>\sim\exp(-E_{max}/\sigma)$ of the jump over the energy barrier for the structural relaxation, which is characterized by an energy barrier distribution with its dispersion $\sigma$ dependent only on the total system entropy $S$ and its maximal energy barrier $E_{max}$ treated as a material constant. Therefore, the Avramov model describes both the density scaling of $\tau$ in terms of Eq. (1) and the structural relaxation time $\tau$ as a single variable function of the total system entropy $S$,

$$\tau = \tau_0 \exp\left[\frac{E_{max}}{\sigma_r}\exp\left(-\frac{2(S-S_r)}{ZR}\right)\right] \quad (2)$$

where $S_r$ and $\sigma_r$ is the total system entropy and the dispersion of the energy barrier distribution in the reference state, $Z$ denotes the number of available pathways for local motions of a molecule or polymer segment, and $R$ is the gas constant. Besides the overall significance of both the relationships (Eqs. (1) and (2)), it should be stressed that the total entropy employed in the Avramov model is more convenient to calculate in comparison with the configurational entropy used in the AG model, which usually requires making some approximations, especially in case of high pressure data. The merits of the Avramov model encourage us to exploit the model as a good example to discuss the suggested dependence $\tau(S)$ in the density scaling regime.

In our comparative study of Eqs. (1) and (2), we use both new and earlier reported experimental dielectric and $pVT$ data of van der Waals liquids such as triethyl-2-acetylcitrate (TBAC), triethyl-2-acetylcitrate (TEAC),[27] 1,1'-bis (p-methoxyphenyl) cyclohexane (BMPC),[28,29] and KDE,[30,31] a polymer polyvinyl acetate (PVAc),[32,33,34] and an ionic liquid 1-butyl-3-methylimidazolium bis(trifluoromethyl-sulfonyl)imide) (BmimNTf2) measured in ambient and high pressure conditions. The isobaric and isothermal dielectric structural relaxation times have been expressed as a function $\tau(T,\rho)$ using a recently derived equation of state for supercooled liquids (Eq. (9) in Ref. 35) with the values of its fitting parameters earlier reported for PVAc,[35] BMPC,[36] KDE,[37] BmimNTf2,[38] TEAC,[27] and determined herein[39] for TBAC. The total system entropy $S$ has been calculated from the well-known thermodynamic formula,

$$S(T,p) = S_r + \int_{T_r}^{T} C_p(T,p_r) d\ln T - \int_{p_r}^{p} \left(\partial V(T,p)/\partial T\right)_p dp \qquad (3)$$

where $S_r=S(T_r,p_r)$ has been determined in the reference state defined by the glass transition temperature at ambient pressure. To calculate $S(T,p)$ from Eq. (3), and then $S(T,V)$, we have exploited the $pVT$ data combined with the isobaric specific heat capacity $C_p$ of TBAC, TEAC, and BmimNTf2 measured herein by differential scanning calorimetry (DSC) with stochastic temperature modulation (TOPEM) at ambient pressure as well as the $C_p$ data earlier reported for BMPC,[36] KDE,[40] and PVAc[41] at ambient pressure.

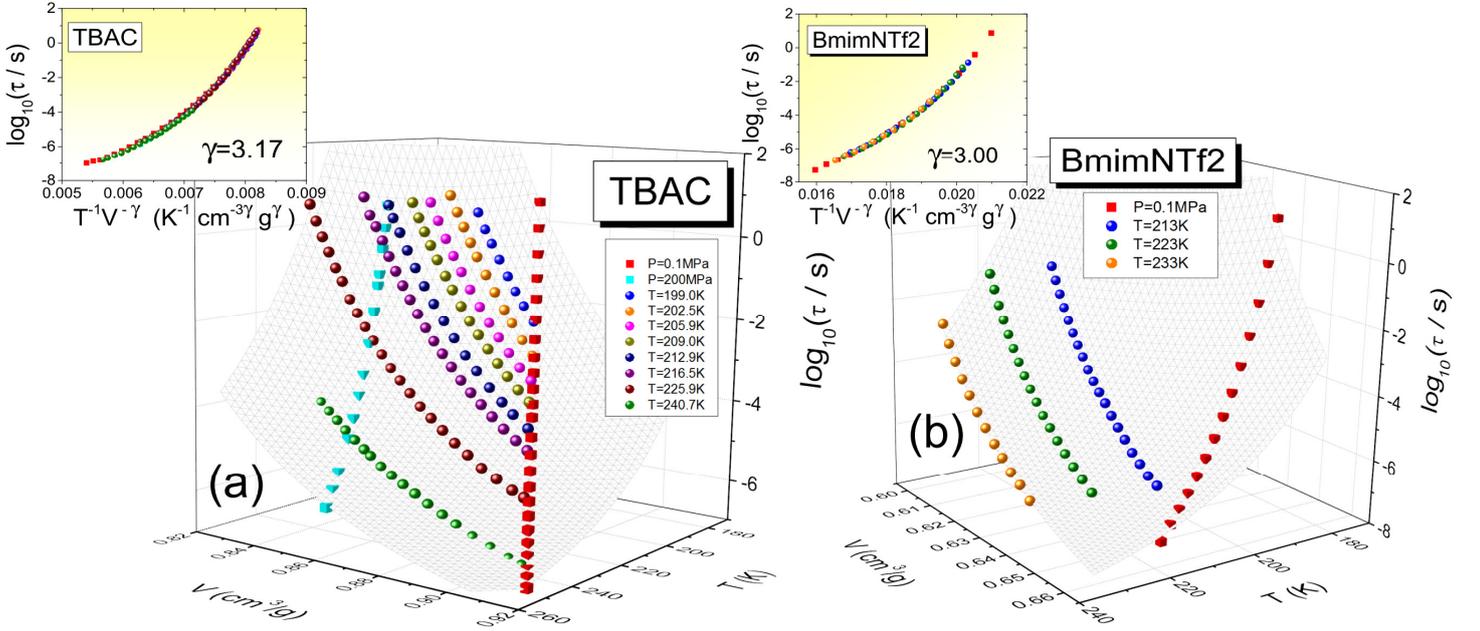

Fig. 1
Plots of the temperature-volume dependences of structural dielectric relaxation times for (a) TBAC and (b) BmimNTf2, respectively. For each material, the fitting surface follows from the fitting of experimental data to Eq. (1), which also yields the value of the scaling exponent $\gamma$ that enables to scale the dependences $\tau(T,V)$ as shown in the inset.

It should be noted that Eq. (1) has been previously very successfully applied to describe the dependences $\tau(T,\rho)$ of many glass formers and the related density scaling $\tau(\rho^\gamma/T)$.[25,40,42,43,44] As examples, we show such results in Fig. 1 for TBAC and BmimNTf2.

However, the latter tempting relation, $\tau(S)$, which suggests that all isobaric and isothermal structural relaxation times can be scaled versus the total system entropy onto a single master curve for a given material, has not been explored thoroughly yet. We perform the test by plotting $\tau$ vs $S$ in the way independent of Eq. (2), i.e., by applying the structural relaxation times $\tau(T,p)$ determined from dielectric spectra and the total system entropy $S(T,p)$ calculated from the thermodynamic equation (Eq. (3)). As a result, we cannot obtain a superposition of the isobaric and isothermal dielectric structural relaxation times for any examined material (see Fig. 2 for TBAC and BmimNTf2 as examples). This test based on experimental data clearly shows a decoupling between isochronal and isoentropic lines, which would never have been observed if $\tau$ had depended only on $S$. As shown in Supporting Information (SI), we have obtained analogous results of this test using the structural relaxation time in the reduced units suggested by the theory of isomorphs to ensure that molecular dynamics is isomorph invariant. It means that the reduced units do not cause the relation between $\tau$ and $S$ to satisfy any single variable function $\tau(S)$.

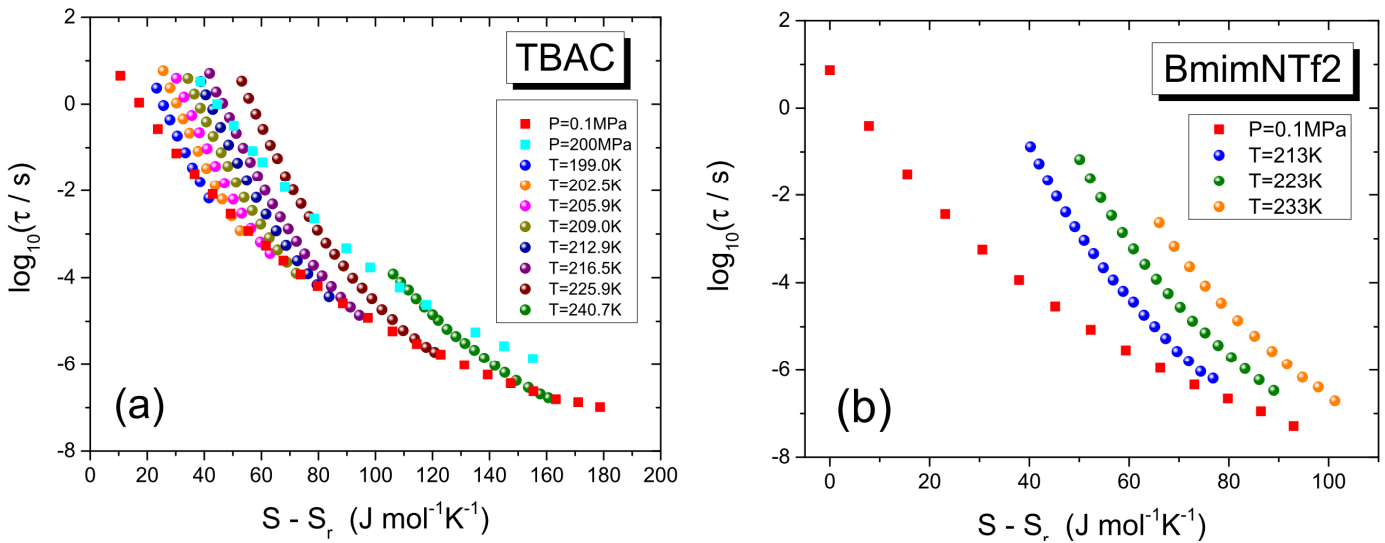

Fig.2

Plots of the structural dielectric relaxation times versus the total system entropy diminished by the total system entropy at a fixed reference state for (a) TBAC and (b) BmimNTf2, respectively.

To answer the question *what can be the reason for the decoupling between τ and S*, we verify whether the total system entropy $S$ can be scaled versus a single variable $\rho^\gamma/T$ or not. It should be noted that the original derivation of Eq. (1) based on Eq. (2), which exploits Eq. (3) with $p$ substituted for $V$ and *vice versa*, involves the following temperature-volume dependence of the total system entropy,[25]

$$S(T,V) = S_r + C_V \ln\left(\frac{TV^{\gamma_S}}{T_r V_r^{\gamma_S}}\right) \qquad (4)$$

where $\gamma_S$ can be expressed by the Grüneisen parameter defined thermodynamically as $\gamma_G \equiv V\alpha_P C_V^{-1} \kappa_T^{-1}$, which involves the volume V, the isochoric specific heat capacity $C_V$, the isobaric thermal volume expansivity $\alpha_p$, and the isothermal compressibility $\kappa_T$. Equation (4) implies that the total system entropy should be scaled versus $\rho^{\gamma_S}/T$, and then $\gamma_S = \gamma$, which can be concluded by comparison of Eqs. (1), (2), and (4). In fact, we have found that $S(T,V)$ can be very well scaled versus $T^{-1}V^{-\gamma_S}$ (see Fig.3 for TBAC and BmimNTf2 as examples, where we use the decimal logarithmic scale for abscissas in the insets to show that the dependence of S-$S_r$ on $\log(T^{-1}V^{-\gamma_S})$ is linear to a good approximation as follows from Eq. (4)). Such results confirm that one can indeed omit a dependence of $C_V$ on $T$ and $V$ in Eq. (4) in the considered experimental range as assumed to formulate this equation.[25] For instance, the mean values of $C_V$ found from the slope of the linear dependence of S-$S_r$ on $\log(T^{-1}V^{-\gamma_S})$ are equal to $C_V$ = (502±1) Jmol$^{-1}$K$^{-1}$ for TBAC and $C_V$ = (404±2) Jmol$^{-1}$K$^{-1}$ for BmimNTf2. However, $\gamma_S \ll \gamma$ and $\gamma_S \approx \gamma_G$ (see Table 1), where γ is established from fitting $\tau(T,V)$ to Eq. (1) and $\gamma_G$ is calculated as a single value for a material from the thermodynamic definition of the Grüneisen parameter in the liquid state near the glass transition at ambient pressure as suggested in Ref. 25. Thus, the decoupling between τ and S can be explained in the density scaling regime in case of each tested material by different values of the scaling exponents $\gamma$ and $\gamma_S$ for $\tau$ and $S$, respectively. It is interesting that we have established (see SI) a similar decoupling between the structural relaxation time $\tau$ and the excess entropy $S_{ex}$, which is the total entropy minus that of an ideal gas at the same density and temperature. This finding is in contrast to some recent results[45,46,47,48,49] based on theoretical and simulation studies, including a standard interpretation[20-22] of the bases of the first version of the theory of isomorphs. This result also suggests that the excess entropy does not give us any advantage over the total system entropy in analyzing the thermodynamic evolution of the time scale of molecular dynamics of real glass formers near the glass transition, because there is in general no single variable function $\tau(S_{ex})$. Nevertheless, our result seems to be in accord with the newly reformulated theory of isomorphs[50,51] (see SI).

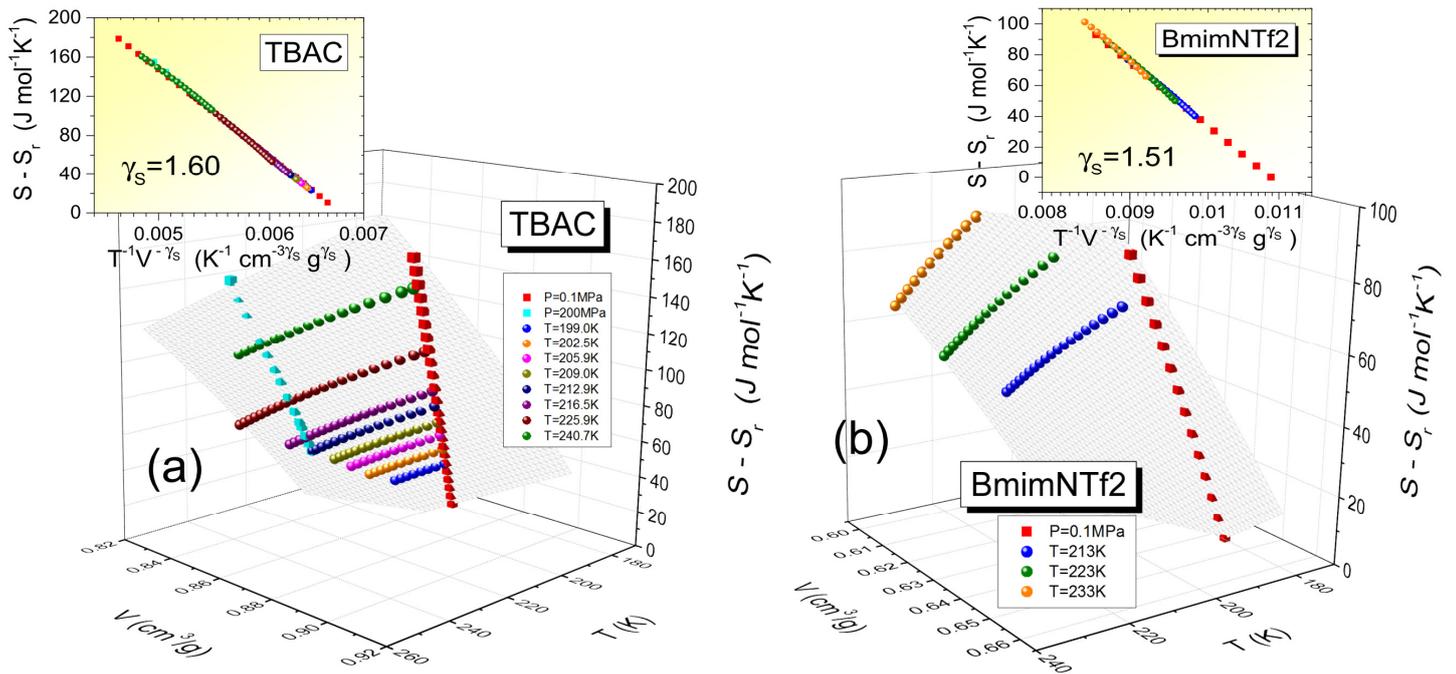

Fig.3

Plots of the temperature-volume dependences of the total system entropy diminished by the total system entropy at a fixed reference state for (a) TBAC and (b) BmimNTf2, respectively. For each material, the fitting surface follows from the fitting of experimental data to Eq. (4), which also yields the value of the scaling exponent $\gamma_S$ that enables to scale the dependences $S-S_r$ $(T,V)$ as shown in the inset, using the decimal logarithmic scale for the inset abscissa.

Very recently, an analogous decoupling has been reported by us[36,52,53] between the structural relaxation time $\tau$ and the novel measure of the dynamic heterogeneity, which is the height of the peak of the four-point dynamic susceptibility, $\chi_4^{max}$. We have shown[53] that if two interrelated quantities obey the density scaling law with different values of the scaling exponents, the interrelation between the quantities requires supplementing with a density power factor. In case of $\tau$ and $S$, the power law density scaling laws $\tau = f(\rho^\gamma/T)$ and $S = h(\rho^{\gamma_S}/T)$ are obeyed if two conditions are met respectively, $\rho^\gamma/T = C_\tau$ at $\tau = const$ and $\rho^{\gamma_S}/T = C_S$ at $S = const$, where $C_\tau$ depends only on $\tau$ and $C_S$ depends only on $S$. Thus, we can establish a criterion for the combined power law density scaling of $\tau$ and $S$, $\rho^{\Delta\gamma} = C_\tau/C_S$ with $\Delta\gamma = \gamma - \gamma_S$, where $C_\tau = const$ and $C_S$ varies with changing density along an isochrone. Then, exploiting the equations, $\tau = f(\rho^\gamma/T)$ and $\rho^{\gamma_S}/T = h^{-1}(S)$, we arrive at a criterion for the combined power law density scaling of $\tau$ and $S$, which defines a non-trivial relation between the time scale of molecular dynamics and the total system entropy,

$$\tau = g\left(\rho^{\Delta\gamma} w(S)\right) \text{ with } \Delta\gamma = \gamma - \gamma_S \qquad (5)$$

where the scaling exponents $\gamma$ and $\gamma_S$ are material constants independent of thermodynamic conditions, and $w = h^{-1}$ is a single variable function of $S$. The criterion given by Eq. (5) can be also generalized to comply with the generalized density scaling laws that allow the density-dependent scaling exponents for $\tau$ and $S$ (as shown in SI).

Thus, the Avramov model (Eqs. (1) and (2)) can be considered in terms of Eq. (5) as an example. Based on Eqs. (1), (2) and (5), one can establish that $\tau = g\left(\rho^{\Delta\gamma} w(S)\right) = \exp\left[\left(\rho^{\Delta\gamma} w(S)\right)^D\right]$, where $D = 2C_V Z^{-1} R^{-1}$ is a parameter of Eq. (1) defined in Ref. 25, while $\rho^{\Delta\gamma} = \gamma - \gamma_S$ follows from Eq. (5) and $w(S) = h^{-1}(S) = \exp(-(S-S_r)/C_V)$ if we assume that Eq. (4) is obeyed, i.e., if $S = h(\rho^{\gamma_S}/T) = S_r - C_V \ln(A_r \rho^{\gamma_S}/T)$, where $A_r = T_r \rho_r^{-\gamma_S}$ and the entropy $S_r$ in the reference state and a mean value of $C_V$ are constant. In this way, we actually postulate that the internal preexponential factor $E_{max}/\sigma_r$ in Eq. (2) depends on a density power. Since $\sigma_r = const$, it physically means that the maximal energy barrier cannot be a material constant and can be scaled via the equation $E_{max} = A_E \rho^{\gamma_E}$ with a scaling exponent $\gamma_E$. We verify this assumption by fitting the dependence $\tau(S,V)$ to Eq. (2) with the parameter $E_{max}$ substituted for its density scaling representation, $A_E \rho^{\gamma_E}$. The obtained values of the scaling exponent $\gamma_E$ for the maximal energy barrier lead to a very satisfactory scaling, $\tau = g(\rho^{\gamma_E} w^D(S))$, where $w^D(S) = (w(S))^D = \exp[-2(S - S_r)Z^{-1}R^{-1}]$ (see Fig. 4). What is more, the density scaling of $E_{max}$ employed in Eq. (2) together with Eq. (4) leads to the same scaling form of Eq. (1), because it results in the relation between the scaling exponents, $\gamma$, $\gamma_S$, and $\gamma_E$,

$$\gamma = \gamma_S + \gamma_E / D \qquad (6)$$

where $D$ is a parameter in Eq. (1), which denotes $D = 2C_V Z^{-1} R^{-1}$ based on Eqs. (2) and (4). Our comparative analysis of Eqs. (1), (2), and (5) clearly shows that Eq. (6) is a consequence of the decoupling between $\tau$ and $S$, the measure of which can be the difference $\Delta\gamma = \gamma - \gamma_S$ between the scaling exponents $\gamma$ and $\gamma_S$. This is because $\gamma_E = D\Delta\gamma$ if $\tau = \exp[(\rho^{\Delta\gamma} w(S))^D]$.

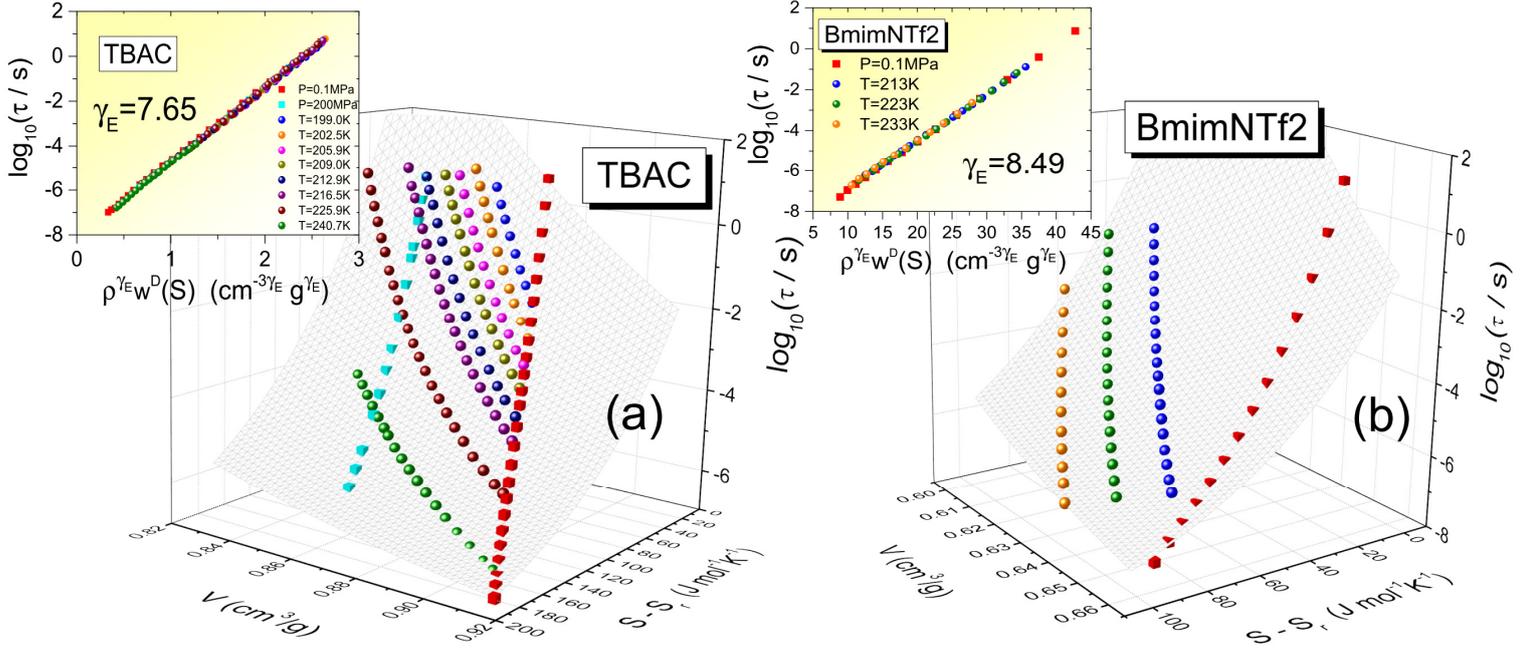

Fig.4

Plots of structural relaxation times versus volume and the total system entropy diminished by the total system entropy at a fixed reference state for (a) TBAC and (b) BmimNTf2, respectively. For each material, the fitting surface follows from the fitting of experimental data to Eq. (2) with $E_{\max} = A_E \rho^{\gamma_E}$, which also yields the value of the scaling exponent $\gamma_E$ that enables to scale the dependences $\tau(V, S-S_r)$ as shown in the inset, where $w^D(S) = \exp[-2(S-S_r)Z^{-1}R^{-1}]$ and $\rho = V^{-1}$.

Table 1

Values of the parameters such as ($\gamma$, $D$), $\gamma_S$, and $\gamma_E$ found from fitting experimental data respectively to Eqs. (1), (4), and (2) with $E_{\max} = A_E \rho^{\gamma_E}$ compared with values of $\gamma_{calc}$ and $\gamma_G$ calculated respectively from Eq. (6) and the thermodynamic definition of the Grüneisen parameter, $\gamma_G \equiv V \alpha_P C_V^{-1} \kappa_T^{-1}$.

| Material | $\gamma_E$ | $\gamma_S$ | $\gamma_G$ | $D$ | $\gamma$ | $\gamma_{calc}$ |
|---|---|---|---|---|---|---|
| TBAC | 7.65±0.12 | 1.60±0.02 | 1.70 | 5.28±0.14 | 3.17±0.03 | 3.05±0.09 |
| TEAC | 7.01±0.14 | 1.74±0.02 | 1.82 | 3.74±0.13 | 3.75±0.02 | 3.62±0.13 |
| BMPC | 11.99±0.18 | 2.12±0.02 | 2.26 | 2.05±0.03 | 8.00±0.09 | 7.97±0.20 |
| KDE | 12.16±0.14 | 0.84±0.03 | 0.86 | 3.09±0.04 | 4.69±0.06 | 4.77±0.13 |
| BmimNTf2 | 8.49±0.19 | 1.51±0.02 | 1.33 | 6.09±0.11 | 3.00±0.05 | 2.91±0.09 |
| PVAc | 6.90±0.12 | 0.74±0.02 | 0.79 | 4.59±0.15 | 2.35±0.03 | 2.24±0.12 |

As can be seen in Table 1, Eq. (6) very well reproduces the values of the scaling exponent $\gamma$ obtained from fitting the structural relaxation times to Eq. (1). The relation has been earlier suggested by us[40] in the form, $\gamma = \gamma_G + \gamma_E / D$, which is equivalent to Eq. (6) due to the correspondence, $\gamma_S \approx \gamma_G$, established herein from the analysis of experimental data. Nevertheless, the crucial physical meaning of Eq. (6) is defined only in this Letter, because this equation directly results from the density scaling of $\tau$ and $S$ with different values of the scaling exponents $\gamma \neq \gamma_S$, and consequently Eq. (6) explicitly shows the contributions of the density scaling of the maximal energy barrier $E_{max}$ and the total system entropy $S$ to the density scaling of the structural relaxation time $\tau$. This clear interpretation is at odds with an earlier less justified attempt[54] to solve the discrepancy $\gamma \neq \gamma_S$ by replacing the total entropy with the configurational entropy in the Avramov model.

The found decoupling between the structural relaxation time $\tau$ and the total system entropy $S$ gives evidence that the entropy cannot only govern the characteristic time scale of molecular dynamics near the glass transition. The additional density factor is required to properly reflect the time scale, $\tau = g(u(\rho)w(S))$. We have established (see SI) almost the same decoupling between $\tau$ and the excess entropy $S_{ex}$, $\tau = \tilde{g}(\tilde{u}(\rho)\tilde{w}(S_{ex}))$, which suggests that the excess entropy should not be favored over the total system entropy in describing the thermodynamic evolution of structural relaxation times. These are similar results to those very recently obtained by us[53] for $\tau$ and the four-point measure of the dynamic heterogeneity $\chi_4^{max}$, for which $\tau = g_\chi(u_\chi(\rho)w_\chi(\chi_4^{max}))$. All the findings clearly show that neither the system entropy (nor the excess entropy) nor the characteristic length scale of the system molecular dynamics are sufficient to represent individually the characteristic time scale of molecular dynamics. The better understanding of the thermodynamic evolution of the structural relaxation times near the glass transition has been achieved in both the cases by exploiting the density scaling idea that has been confirmed to be able to separate physically relevant contributions to the time scale of molecular dynamics.

**Acknowledgements**

The authors thank a Ph. D. student G. Jarosz for providing dielectric data of BmimNTf2. K.G., A.G., and M.P. are grateful for the financial support from the Polish National Science Center based on Decision No. DEC-2012/04/A/ST3/00337 within the program MAESTRO 2.

**Supporting Information**

The Supporting Information includes: (S.1) an analysis of the density scaling of the total system entropy in reduced units and in terms of different reference states, (S.2) a comparison of the total system entropy and the excess entropy in describing the thermodynamic evolution of structural relaxation times, and (S.3) the combined density scaling law for $\tau$ and $S$ in the general case of the density scaling.

# Role of entropy in the thermodynamic evolution of the time scale of molecular dynamics near the glass transition

K. Grzybowska,[1,2,*] A. Grzybowski,[1,2,#] S. Pawlus,[1,2] J. Pionteck,[3] and M. Paluch[1,2]

[1]*Institute of Physics, University of Silesia, Uniwersytecka 4, 40-007 Katowice, Poland*

[2]*Silesian Center for Education and Interdisciplinary Research, 75 Pułku Piechoty 1A, 41-500 Chorzów, Poland*

[3]*Leibniz Institute of Polymer Research Dresden, Hohe Str. 6, D-01069 Dresden, Germany*

[*] Corresponding author's email: katarzyna.grzybowska@us.edu.pl

[#] Corresponding author's email: andrzej.grzybowski@us.edu.pl

## SUPPORTING INFORMATION

### S.1 Analysis of the density scaling of the total system entropy in reduced units and in terms of different reference states

As mentioned in the main part of the Letter, a solid theoretical ground for the density scaling idea has been formulated by Dyre's group in the theory of isomorphs.[20] According to this theory, the density scaling laws should be validated by using some reduced units for different physical quantities to ensure that molecular dynamics is isomorph invariant. Although the reduced units ensure that molecular dynamics is invariant only on assumption NVT or NVE statistical ensembles while experimental data are measured rather in the conditions of the NPT statistical ensemble, it is worth verifying the density scaling laws for the structural relaxation time and the total system entropy, $\tau = f(\rho^\gamma/T)$ and $S = h(\rho^{\gamma_S}/T)$, by using structural relaxation times in the reduced units defined as follows $\tilde{\tau} = \tau \rho^{1/3} (k_B T / M)^{1/2}$, where $M$ is the average particle mass (or the average mass of polymer unit if $\tau$ is the segmental relaxation time of polymer) and $k_B$ is the Boltzmann constant. As an example, we present herein such an analysis for 1,1'-bis (p-methoxyphenyl) cyclohexane (BMPC), which is a typical van der Waals liquid. First, in Fig. S.1, we compare the density scaling of structural relaxation times $\tau$ and $\tilde{\tau}$, and then we plot the dependences $\tau(S)$ and $\tilde{\tau}(S)$ in Fig. S.2, showing that the reduced units suggested by the theory of isomorphs do not improve the relation between the structural relaxation time and the total system entropy, because the curves in Fig. S.2(b) do not superimpose similarly to those in Fig. S.2(a).

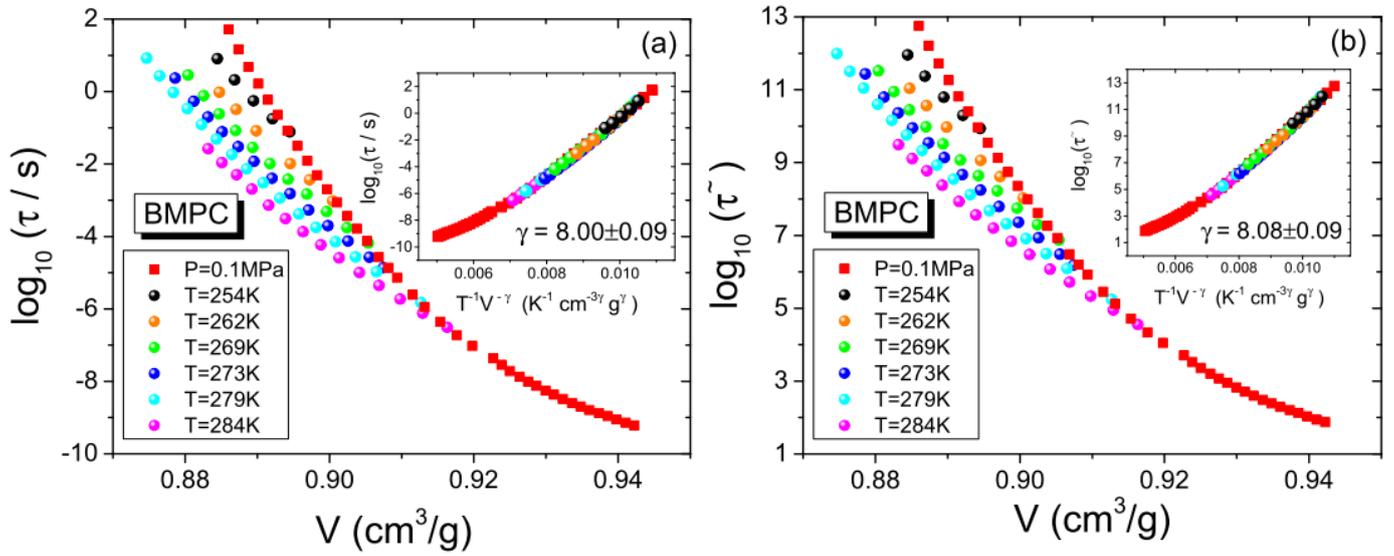

Fig. S.1
Plots of the volume dependences of structural relaxation times $\tau$ and $\tilde{\tau}$ respectively in (a) seconds and (b) reduced units suggested by the theory of isomorphs. The insets show the power law density scaling of $\tau$ and $\tilde{\tau}$.

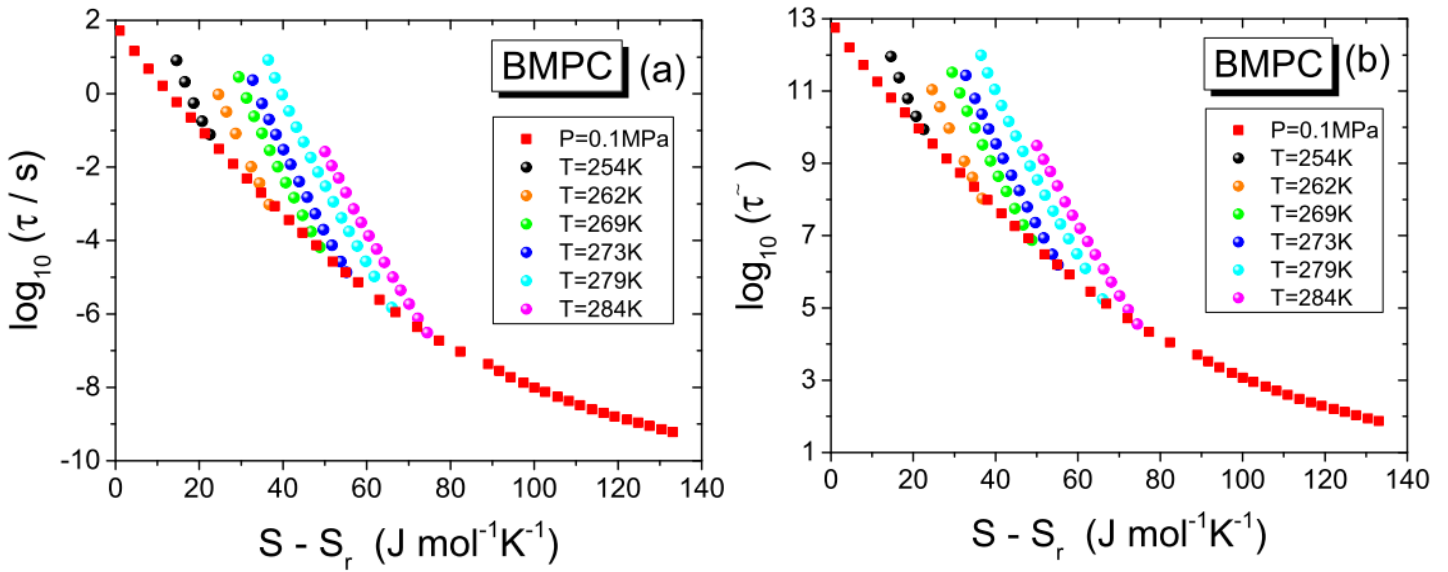

Fig. S.2
Plots of structural relaxation times $\tau$ and $\tilde{\tau}$ respectively in (a) seconds and (b) reduced units suggested by the theory of isomorphs versus the total system entropy $S$ calculated relative to the total system entropy $S_r$ in the reference state chosen at the glass transition temperature at ambient pressure.

In addition, to avoid any doubts about any influence of the chosen reference state to calculate the total system entropy on our analyses, we positively verify the analyses (see Fig. S.3) by repeating calculations using the melting temperature at ambient pressure, $T_{m_0} = T_m(p_0)$ where $p_0=0.1$MPa, as a reference state $(T_{m_0}, p_0)$ instead of $(T_{g_0}, p_0)$, where $T_{g_0} = T_g(p_0)$ is the glass transition temperature at ambient pressure, which has been assumed to perform all previous calculations.

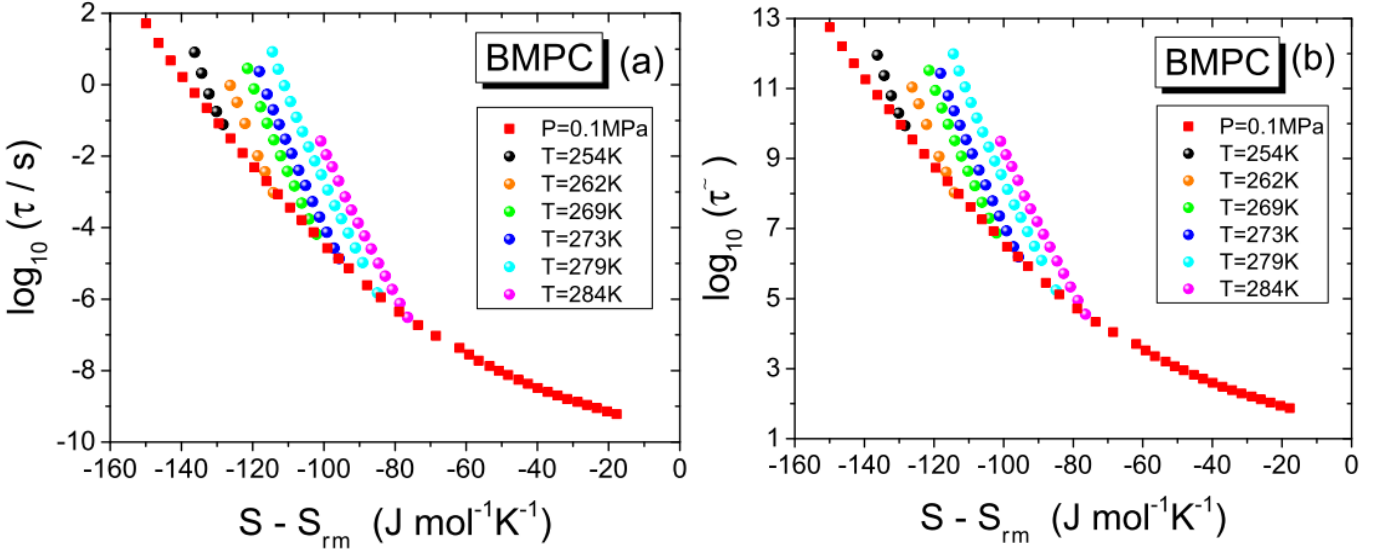

Fig. S.3
Plots of structural relaxation times $\tau$ and $\tilde{\tau}$ respectively in (a) seconds and (b) reduced units suggested by the theory of isomorphs versus the total system entropy S calculated relative to the total system entropy $S_r$ in the reference state chosen at the melting temperature at ambient pressure.

### S.2 Comparison of the total system entropy and the excess entropy in describing the thermodynamic evolution of structural relaxation times

Starting from Rosenfeld's works on the role of entropy in transport properties of simple fluids,[45] one can expect that some dynamic quantities as viscosity, diffusivity, and structural relaxation time should be scaled with the excess entropy $S_{ex}$ defined as the total system entropy diminished by that of an ideal gas at the same density and temperature. In the first version of the isomorph theory, the excess entropy plays a special role as a fundamental quantity, the value of which is invariant along an isomorph. Based on the temperature factorization, $k_B T = f(S_{ex})\hat{h}(\rho) = const$, the first version of the isomorph theory determines a criterion for the generalized density scaling law, $\hat{h}(\rho)/T = const$ at $S_{ex} = const$, which is met along an isomorph, where any single variable function $f(S_{ex}) = const$, because $S_{ex} = const$ along an isomorph. As a consequence of the temperature factorization, the scaling exponent has been defined as $\gamma \equiv (\partial \ln T / \partial \ln \rho)_{S_{ex}}$ and concluded as a function, $\gamma(\rho) = d \ln \hat{h}(\rho) / d \ln \rho$, which can be in general dependent on density. As can be easily seen, this formalism leads to $\gamma = const$ in the limit of the power law density scaling, i.e., if $\hat{h}(\rho) = \rho^{\gamma}$ with γ treated as a material constant independent of thermodynamic conditions. Considering this case, we have verified whether or not the structural relaxation time τ is a single variable function of $S_{ex}$ as suggested by the first version of the isomorph theory, according to which the structural relaxation time in the reduced units $\tilde{\tau}$ is invariant along an isomorph similarly as $S_{ex}$. In other words, since both $\tilde{\tau}$ and $S_{ex}$ are invariant along an isomorph, both $\tilde{\tau}$ and $S_{ex}$ should be scaled with the same value of the scaling exponent γ at least to a good approximation. Continuing the analysis for BMPC as an example, we have found that neither τ nor $\tilde{\tau}$ is any single variable function of $S_{ex}$, because the curves in both the panels of Fig. S.4 do not superimpose. Therefore, the use of the excess entropy in describing the thermodynamic evolution of the time scale of molecular dynamics near the glass transition do not give any advantage over the total system entropy. What is more, we have established that the excess entropy obeys a power law density scaling law (the inset in Fig. S.5(c)) with the scaling exponent $\gamma_{S_{ex}}$, the value of which considerably differs from that which scales structural relaxation times γ and is pretty close to that which scales the total system entropy $\gamma_S$.

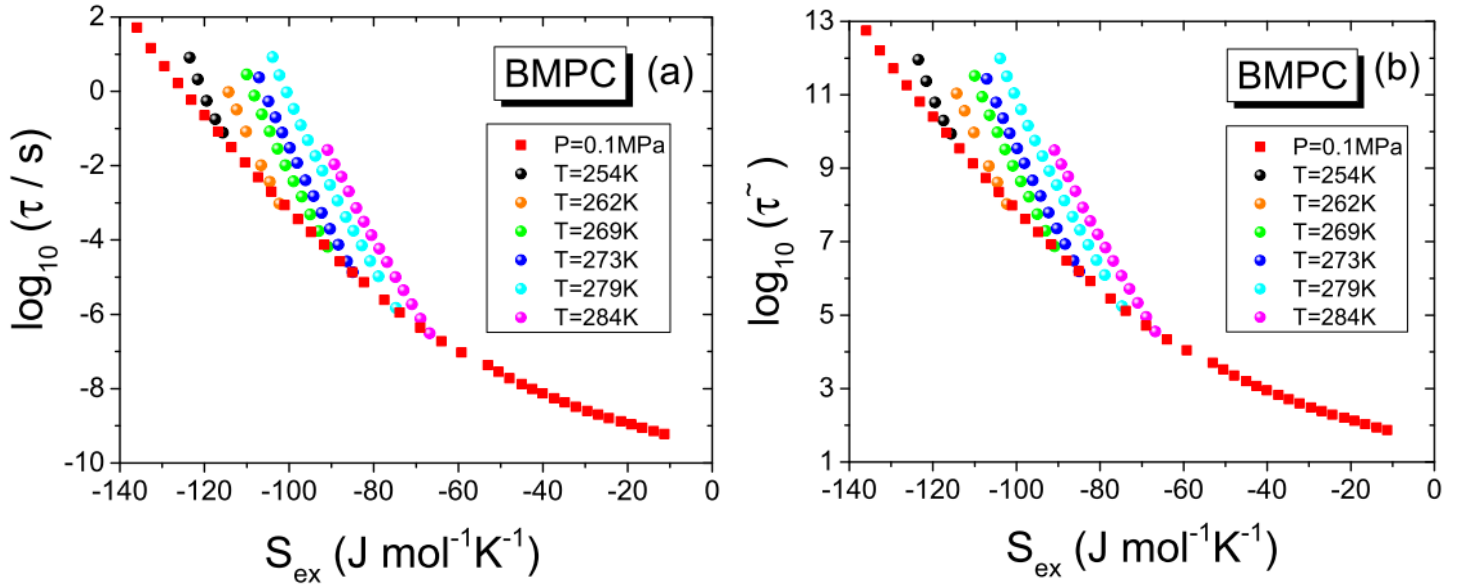

Fig. S.4
Plots of structural relaxation times τ and $\tilde{\tau}$ respectively in (a) seconds and (b) reduced units suggested by the theory of isomorphs versus the excess system entropy $S_{ex}$, which is in excess of that of an ideal gas at the same temperature and density.

The latter finding suggests that the excess entropy can comply with a density scaling law similar to that valid for the total system entropy. It is actually reasonable because Eq. (4), given in the main part of the Letter, can be easily modified to that valid for excess quantities,

$$S_{ex}(T,V) \approx S_r^{ex} + C_V^{ex} \ln\left(\frac{TV^{\gamma_{S_{ex}}}}{T_r V_r^{\gamma_{S_{ex}}}}\right) \tag{S.1}$$

where $S_r^{ex} = S_{ex}(T_r, V_r)$ and $\gamma_{S_{ex}}$ can be expressed by the excess Grüneisen parameter defined thermodynamically as $\gamma_G^{ex} \equiv V\alpha_P C_V^{ex^{-1}} \kappa_T^{-1}$ with the excess isochoric heat capacity $C_V^{ex}$ calculated as $C_V^{ex} = C_V - C_V^{ideal\ gas}$. It should be noted that Eq. (S.1) is an approximate equation because of two assumptions: (i) $C_V^{ex} \approx const$ and $\gamma_G^{ex} \approx const$, by analogy with Eq. (4) given in the main part of the Letter, (ii) and additionally Eq. (S.1) is simplified by neglecting a density dependent term, $-R\ln(V/V_r)$, the value of which is nearly constant and small. For instance, in case of BPMC, the absolute value of this term does not exceed 0.5Jmol$^{-1}$K$^{-1}$, while the absolute value of the calculated excess entropy $|S_{ex}|$ ranges from 10Jmol$^{-1}$K$^{-1}$ to 140 Jmol$^{-1}$K$^{-1}$. We have tested Eq. (S.1). As an example, we demonstrate the density scaling plot of $S_{ex}$ vs $T^{-1}V^{-\gamma_{S_{ex}}}$, where we use the decimal logarithmic scale for abscissa in the inset to show that the dependence of $S_{ex}$ on $\log(T^{-1}V^{-\gamma_{S_{ex}}})$ is linear to a good approximation as follows from Eq. (S.1). We have calculated the excess entropy of BMPC as the total entropy of this glass-forming liquid minus that of an polyatomic ideal gas at the same density-temperature conditions to perform the analysis shown in Figs. S.4 and S.5(c). It means that $C_V^{ideal\ gas} = 3R$, where the gas constant $R \approx 8.31446$ Jmol$^{-1}$K$^{-1}$. Thus, if Eq. (S.1) is obeyed, the following equation should be met, $C_V - C_V^{ex} = C_V^{ideal\ gas}$, where the mean values of $C_V$ and $C_V^{ex}$ are obtained respectively by fitting the evaluated dependences of $S - S_r$ on $\log(T^{-1}V^{-\gamma_{S_{ex}}})$ and $S_{ex}$ on $\log(T^{-1}V^{-\gamma_{S_{ex}}})$ to a linear function according respectively to Eq. (4) in the main part of the Letter and Eq. (S.1). As an example, the fitting values obtained for BMPC, $C_V = (323\pm1)$ Jmol$^{-1}$K$^{-1}$ and $C_V^{ex} = (298\pm1)$ Jmol$^{-1}$K$^{-1}$, have turned out to satisfy very well the expected relation $C_V - C_V^{ex} = 3R$, because $C_V - C_V^{ex} = (25\pm1)$ Jmol$^{-1}$K$^{-1}$ and $3R \approx 24.94$ Jmol$^{-1}$K$^{-1}$. In this way, the combined test for the validity of Eq. (4) in the main part of the Letter and Eq, (S.1) has been very satisfactorily passed. It should be stressed that such results of the test successfully validate our analyses of the density scaling laws for $S$ and $S_{ex}$.

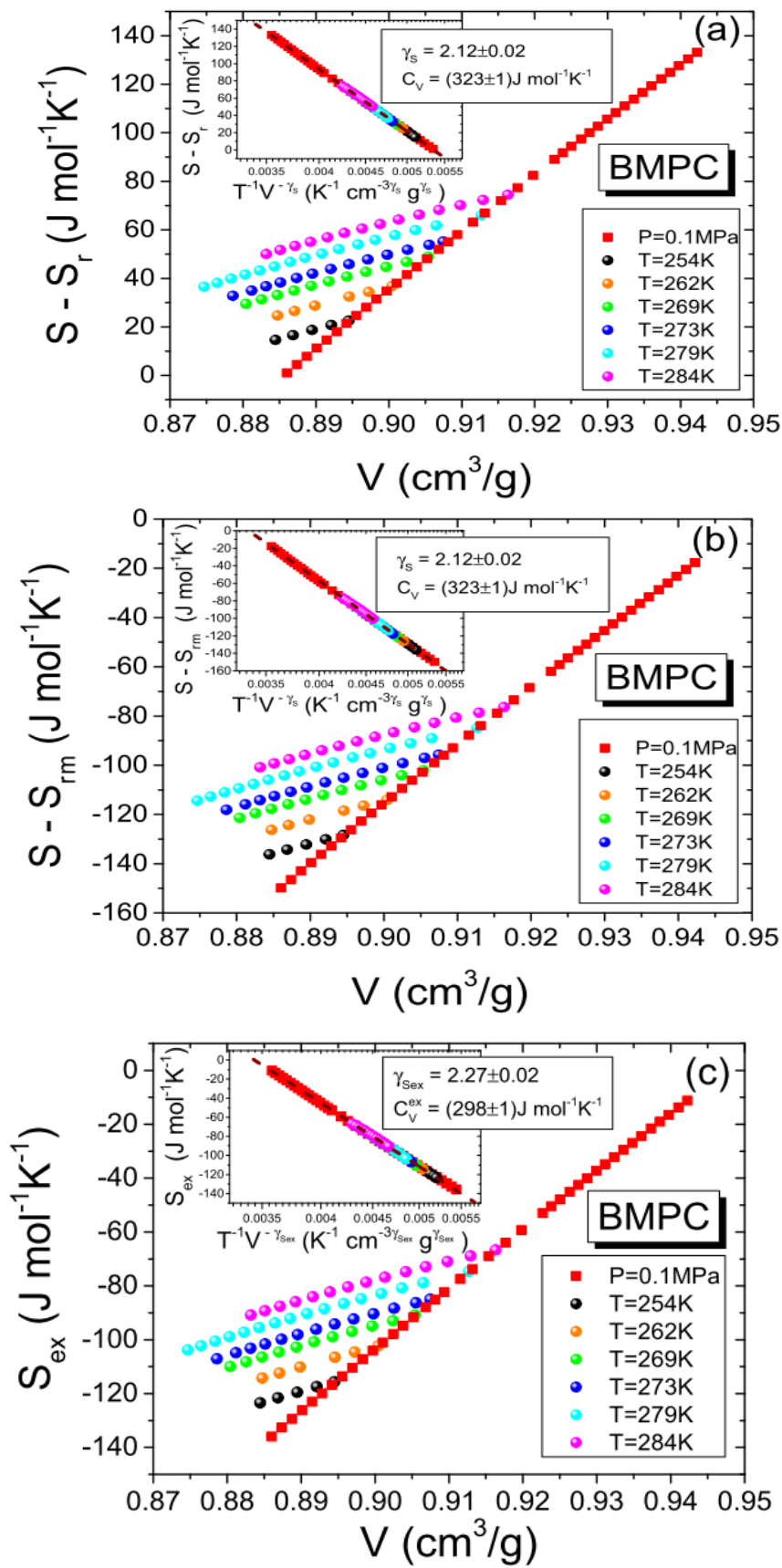

Fig. S.5

Plots of the volume dependences of the total system entropy calculated relative to the reference total system entropy chosen at ambient pressure (a) at the glass transition temperature and (b) at the melting temperature, and (c) the excess entropy calculated in excess of that of an ideal gas at the same temperature and density. The insets represent the corresponding density scaling laws, using the decimal logarithmic scales for their abscissas.

An important conclusion drawn from the density scaling law for $S_{ex}$, which is obeyed by real glass formers with a value of the scaling exponent $\gamma_{S_{ex}}$ that considerably differs from the scaling exponent $\gamma$ that scales structural relaxation times $\tau$, are different patterns of isomorphs for $\tau$ and $S_{ex}$, which have not been expected by the authors of the first version of the isomorph theory. This is because we have shown that both the density scaling laws for $\tau$ and $S_{ex}$ are obeyed but with different values of the scaling exponents $\gamma$ and $\gamma_{S_{ex}}$. It means that the following criteria for the density scaling laws are met, $\hat{h}(\rho)/T = C_{S_{ex}}$ at $S_{ex} = const$ and $\tilde{h}(\rho)/T = C_\tau$ at $\tau = const$, where $C_{S_{ex}}$ depends only on $S_{ex}$ and $C_\tau$ depends only on $\tau$. Consequently, the density-dependent scaling functions $\hat{h}(\rho)$ and $\tilde{h}(\rho)$ are in general different and the density scaling criteria for $\tau$ and $S_{ex}$ define different isomorphic curves in the phase diagram. As an example, in Fig. S.6, we illustrate results of such tests based on the criteria in the limit of the power law density scaling, which indeed show that the values of the scaling exponent for $\tau$ (or $\tilde{\tau}$) and $S$ (or $S_{ex}$) are considerably different. In our opinion, such a decoupling between isomorphs for $\tau$ and $S_{ex}$ could be alternatively considered within the framework of the first version of the isomorph theory if this theory resigned itself to maintain that the definition of the scaling exponent $\gamma \equiv (\partial \ln T / \partial \ln \rho)_{S_{ex}}$ could be transferred to $\gamma = (\partial \ln T / \partial \ln \rho)_{\tilde{\tau}}$ to a good approximation. This is certainly not any standard interpretation of the bases of the first version of the isomorph theory. Nevertheless, very recent developments of the isomorph theory, which tend to achieve a better accordance of the isomorph theory predictions with the properties established experimentally for real materials, remove some previous limitations of the theory. Bohling et al.[50] have very recently argued that the isomorph criterion can be extended to the form $\tilde{h}(\rho, S_{ex})/T = const$, which should better reflect the density scaling properties of real glass formers that are more complex than simple model systems previously used to test the isomorph theory. The extended density scaling criterion has found its theoretical grounds in the newly reformulated theory of isomorphs combined with a general definition of Roskilde-simple systems.[51] Our findings obtained by using experimental data clearly show that there is a decoupling between $\tau$ and $S$ as well as between $\tau$ and $S_{ex}$, which can be in general described as follows, $\tau = g(u(\rho)w(S))$ and $\tau = \tilde{g}(\tilde{u}(\rho)\tilde{w}(S_{ex}))$, respectively. Considering the structural relaxation time $\tau$ as invariant along each isomorph determined from the new criterion $\tilde{h}(\rho, S_{ex})/T = const$, we can draw the same conclusion in relation to the excess entropy that there is in general no single variable function $\tau(S_{ex})$. However, we can successfully use also the total system entropy instead of the excess entropy in describing the thermodynamic evolution of the time scale of molecular dynamics near the glass transition, which certainly simplifies further theoretical investigations in this field.

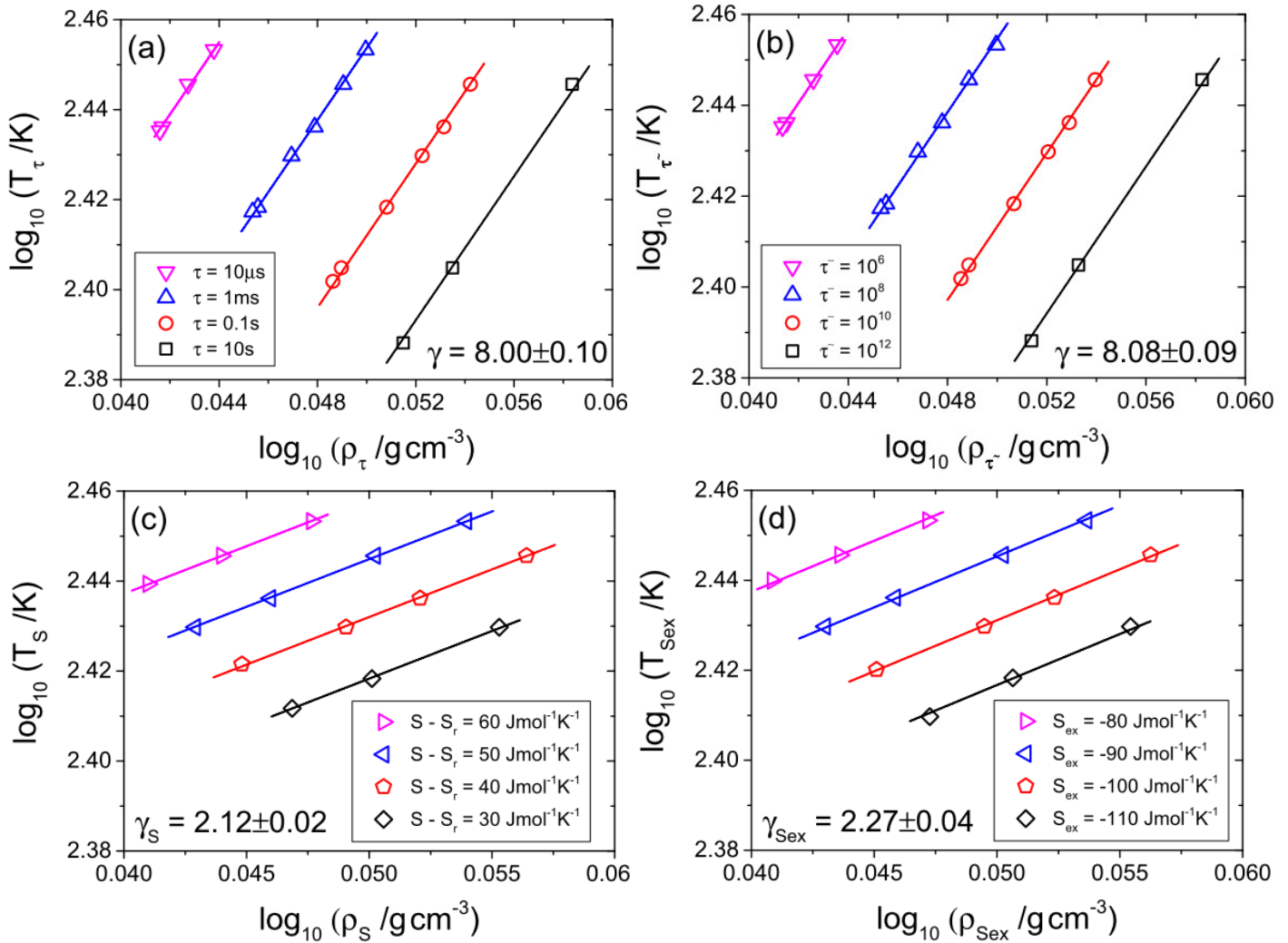

Fig. S.6
Double logarithmic plots of temperature versus density established along a few (a) isochrones defined at constant structural relaxation times in seconds, (b) isochrones defined at constant structural relaxation times in the reduced units suggested by the theory of isomorphs, (c) isoentropic curves defined at constant total system entropies (d) isoentropic curves defined at constant excess entropies which is in excess of that of an ideal gas at the same temperature and density. The values of the scaling exponents $\gamma$, $\gamma_S$, and $\gamma_{Sex}$ are found as slopes of the linear fits denoted by solid lines.

## S.3 The combined density scaling law for $\tau$ and $S$ in the general case of the density scaling

In the general case of the density scaling, any constant exponent γ does not enable us to achieve the power law density scaling of τ, and to do that, we need to employ other usually more complex functions, which depends (at least) only on density.[21,22,50,51] It is reasonable to assume that the same problem concerns the density scaling of $S$ and $S_{ex}$ in the extremely wide pressure range. Then, the criterion $\rho^{\Delta\gamma} = C_\tau / C_S$ for the combined density scaling of τ and $S$ should be extended to $\Re(\rho) = C_\tau / C_S$ with a density function, $\Re(\rho) = \Im_\tau(\rho)/\Im_S(\rho)$, where $\Im_\tau(\rho)/T$ and $\Im_S(\rho)/T$ are respectively the density scaling variables for τ and S, where an analogous pattern of the density scaling behavior can be drawn for τ and $S_{ex}$. Thus, Eq. 5 from the main part of the Letter can be generalized as follows,

$$\tau = g\big(\Re(\rho)w(S)\big) \text{ with } \Re(\rho) = \Im_\tau(\rho)/\Im_S(\rho), \tag{S.2}$$

if the density scaling laws, $\tau\big(\Im_\tau(\rho)/T\big)$ and $S\big(\Im_S(\rho)/T\big)$, are obeyed, where we can alternatively consider also $S_{ex}\big(\Im_{S_{ex}}(\rho)/T\big)$. Nevertheless, we have limited all the tests presented in this Letter to the case of the simple scaling variables, $\rho^\gamma/T$ and $\rho^{\gamma_S}/T$, including $\rho^{\gamma_{Sex}}/T$, because experimental data that comply with the power law density scaling rule are commonly accessible in contrast to those measured at very high pressures sufficient to reveal the more complex density scaling behavior. As an example, using experimental data of BMPC, it has been shown that our experimental data are limited to the power law density scaling regime, because all double logarithmic dependences are linear to a very good approximation in all panels of Fig. S.6. What is more, the slopes of the appropriate isochronal and isoentropic dependences in individual panels of Fig. S.6 are exactly the same as the scaling exponents which enable to scale respectively structural relaxation times τ and $\tilde{\tau}$ in seconds and the reduced units suggested by the theory of isomorphs (as shown in the insets in Figs. S.1 (a) and (b)) as well as the total system entropy $S$ (scaled in the insets in Figs. S.5 (a) and (b)) and the excess entropy $S_{ex}$ (scaled in the inset in Fig. S.5 (c)).